\documentclass[journal]{new-aiaa}
\usepackage[utf8]{inputenc}

\usepackage{amsmath,amssymb}
\usepackage{graphicx}
\usepackage[version=4]{mhchem}
\usepackage{booktabs}
\usepackage{siunitx}

\title{Spectral Reconstruction for Under-Resolved Turbulence Measurements Using a Variational Cutoff Dissipation Model}

\author{Rishabh Mishra\footnote{Research Fellow,  Laboratoire de recherche en Hydrodynamique, Énergétique et Environnement Atmosphérique,
UMR CNRS 6598; rishabh.mishra@ec-nantes.fr.}}
\affil{LHEEA lab, CNRS -- Centrale Nantes, Nantes Universit\'{e}, 44100 Nantes, France}

\begin{document}

\maketitle

\noindent
\begin{tabular}{@{}l @{\quad=\quad} l}
$C_K$      & Kolmogorov constant ($\approx 1.5$) \\
$E(k)$     & turbulence energy spectrum, m$^3$/s$^2$ \\
$f_c$      & cutoff frequency of the sensor, Hz\\
$k$        & wavenumber, m$^{-1}$ \\
$k_d$      & dissipation onset wavenumber, m$^{-1}$ \\
$k_\eta$   & Kolmogorov wavenumber ($= 1/\eta$), m$^{-1}$ \\
$k_{max}$ & maximum wavenumber of a truncated spectrum, m$^{-1}$ \\
$Re_\lambda$ & Taylor microscale Reynolds number \\
$\mathcal{T}(\kappa)$     & spectral transmission coefficient \\
$Z$        & conformal resistance coordinate \\[6pt]
\multicolumn{2}{@{}l}{\textit{Greek Symbols}} \\[3pt]
$\kappa$ & normalised wavenumber defined as $\kappa = k \eta$\\
$\Gamma(\kappa)$ & cascade resistance field \\
$\varepsilon$ & turbulent energy dissipation rate, m$^2$/s$^3$ \\
$\eta$     & Kolmogorov length scale ($= (\nu^3/\varepsilon)^{1/4}$), m \\
$\lambda$ & Taylor microscale, m \\
$\nu$      & kinematic viscosity, m$^2$/s \\
\end{tabular}

\section{Introduction}

\noindent Accurate characterization of turbulence intensity and turbulent kinetic energy (TKE) is essential for aerodynamic design, combustion modeling, and environmental flow assessment \cite{Pope2000,Tennekes1972}. The standard approach involves measuring the turbulence energy spectrum $E(k)$ and integrating over all wavenumbers $k$:
\begin{equation}
    \text{TKE} = \int_0^\infty E(k)\, dk
\end{equation}

\noindent In practice, this integration requires resolving the spectrum across all energetically significant scales---from the energy-containing range through the dissipation range near the Kolmogorov wavenumber $k_\eta = 1/\eta$, where $\eta = (\nu^3/\varepsilon)^{1/4}$ is the Kolmogorov length scale \cite{Kolmogorov1941}.

\subsection{The resolution problem}

\noindent Hot-wire anemometry remains the gold standard for turbulence measurements due to its high frequency response \cite{Bruun1995,Comte-Bellot1976}. However, hot-wires present significant practical challenges that limit their applicability in many situations. These include fragility in contaminated or particle-laden flows \cite{Fingerson1994}, stringent calibration requirements that must be repeated frequently to account for probe drift \cite{Jorgensen2002}, sensitivity to temperature variations that necessitate compensation or controlled environments \cite{Bruun1995}, and the high cost and specialized expertise required for proper operation \cite{Tropea2007}.

\noindent Pressure-based sensors, such as dynamic pressure transducers, microphones, and piezoresistive sensors (e.g., Kulite transducers), offer significant advantages in terms of robustness and ease of deployment \cite{Blake2017,Willmarth1975}. However, these sensors typically suffer from bandwidth limitations that prevent resolution of the high-wavenumber portion of the turbulence spectrum. A sensor with cutoff frequency $f_c$ will truncate the measured spectrum at:
\begin{equation}
    k_{max} = \frac{2\pi f_c}{U}
\end{equation}
where $U$ is the mean convection velocity. When $k_{max} < k_\eta$, the dissipation range remains unresolved.

\subsection{Consequences of spectral truncation}

\noindent Integrating a truncated spectrum yields systematic underestimation of TKE. While the dissipation range contains relatively little energy compared to the inertial and energy-containing ranges, the deficit is not negligible. Analysis of high-Reynolds-number spectra \cite{Saddoughi1994,Pope2000} indicates that for a Kolmogorov spectrum truncated at $k\eta = 0.2$, the missing energy fraction is approximately 8--12\% depending on the Reynolds number and the extent of the inertial subrange.

\noindent Moreover, accurate estimation of the dissipation rate $\varepsilon$ requires resolving the spectral rolloff in the dissipation range \cite{Antonia1991,Sreenivasan1998}. When this rolloff is not captured, conventional methods for determining $\varepsilon$ from the spectrum become unreliable.

\subsection{Existing dissipation range models}

\noindent Classical spectral models extend the Kolmogorov inertial-range spectrum \cite{Kolmogorov1941,Obukhov1941} into the dissipation range using exponential decay functions. The model of Pao \cite{Pao1965} assumes that the ratio of energy flux to energy is independent of viscosity, yielding:
\begin{equation}
    E(k) = C_K \varepsilon^{2/3} k^{-5/3} \exp\left(-\frac{3}{2} C_K (k\eta)^{4/3}\right)
    \label{eq:pao}
\end{equation}

\noindent Pope \cite{Pope2000} proposed a more general form with adjustable parameters:
\begin{equation}
    E(k) = C_K \varepsilon^{2/3} k^{-5/3} f_\eta(k\eta)
    \label{eq:pope}
\end{equation}
\noindent where $f_\eta(k\eta) = \exp\left\{-\beta\left([(k\eta)^4 + c_\eta^4]^{1/4} - c_\eta\right)\right\}$ with recommended values $\beta \approx 5.2$ and $c_\eta \approx 0.4$ based on experimental fits \cite{Pope2000,Saddoughi1994}. It should be noted that Pope's general model includes terms for both the energy-containing and dissipation ranges. Since this study focuses only on the dissipative part of the spectrum, only the part which defines the dissipative part ($f(k\eta)$) is used. It is reasonable to do so because the function describing the energy-containing part of the spectrum tends to unity in dissipative part of the spectrum. The full spectral equation is available in section 6.5.3 of Pope \cite{Pope2000}.

\noindent While these models capture the spectral decay reasonably well, they predict exponential tails extending to $k \to \infty$. This presents difficulties for spectral integration: any numerical evaluation requires arbitrary truncation, and the fitted parameters exhibit Reynolds-number dependence \cite{Martinez1997,Ishihara2005}.

\subsection{Present contribution}

\noindent This paper presents an analytical model for the dissipation range that enables reconstruction of the full spectrum from partial measurements where the inertial subrange has been resolved at least partially. The proposed model features:
\begin{enumerate}
    \item A cutoff at the Kolmogorov wavenumber ($E(k_\eta) = 0$ exactly)
    \item No adjustable parameters beyond the Kolmogorov constant $C_K$
    \item Bounded spectral support enabling unambiguous integration
\end{enumerate}

\noindent The model is derived from a variational principle governing the cascade resistance in wavenumber space, yielding a Ginzburg-Landau domain wall solution \cite{Goldenfeld1992}. Experimental validation demonstrates that the proposed model accurately captures the shape of the dissipative part of the energy spectrum. Furthermore, it achieves superior accuracy in TKE recovery compared to the formulations of Pao \cite{Pao1965} and Pope \cite{Pope2000}.

\section{Theoretical formulation}

\subsection{Spectral model structure}

\noindent The energy spectrum is modeled as the product of the Kolmogorov inertial-range form and a transmission function:
\begin{equation}
    E(k) = C_K \varepsilon^{2/3} k^{-5/3} \mathcal{T}(k\eta)
    \label{eq:spectrum}
\end{equation}
\noindent where $C_K \approx 1.5$ is the Kolmogorov constant \cite{Sreenivasan1995} and $\mathcal{T}(\kappa)$ is a dimensionless transmission function with normalised wave number $\kappa = k\eta$.

\noindent The transmission function must satisfy the boundary conditions:
\begin{itemize}
    \item $\mathcal{T}(\kappa) \to 1$ as $\kappa \to 0$ (inertial range recovery)
    \item $\mathcal{T}(\kappa) \to 0$ as $\kappa \to 1$ (dissipation cutoff)
\end{itemize}

\subsection{Cascade resistance and transmission}

\noindent The cascade resistance $\Gamma(\kappa) \in [0,1]$ is defined as the fraction of energy flux that has been removed from the cascade by viscous dissipation up to normalised wavenumber $\kappa$:
\begin{equation}
    \Gamma(\kappa) \equiv 1 - \frac{\Pi(\kappa)}{\varepsilon}
\end{equation}
\noindent where $\Pi(\kappa)$ is the spectral energy flux through non-dimensional wavenumber $\kappa$ \cite{Pope2000}. In the inertial range, $\Pi = \varepsilon$ and thus $\Gamma = 0$; at the Kolmogorov scale, $\Pi = 0$ and $\Gamma = 1$.

\noindent Following the analysis of nonlinear energy transfer in turbulence \cite{Kraichnan1959,Zhou2021}, the spectral transmission is related to the cascade resistance by:
\begin{equation}
    \mathcal{T}(\kappa) = (1 - \Gamma^2)^2
    \label{eq:transmission_def}
\end{equation}

\noindent This quartic form arises from the structure of triad interactions in the energy cascade, where the transfer rate depends on the product of spectral amplitudes at interacting wavenumbers \cite{Domaradzki1990,Eyink2005}.





\subsection{Variational principle for cascade dynamics}

\noindent To determine the functional form of $\Gamma(\kappa)$, a variational approach is employed. The dissipation range spans the finite interval $\kappa \in [\kappa_d, 1]$, where $\kappa_d = k_d \eta$ marks the onset of viscous effects. A conformal coordinate is introduced:
\begin{equation}
    Z = \sqrt{\frac{\kappa - \kappa_d}{1 - \kappa}}
    \label{eq:conformal}
\end{equation}
\noindent which maps $[\kappa_d, 1) \rightarrow [0, \infty)$, regularizing the boundary condition at the Kolmogorov scale.

\begin{figure}[hbt!]
    \centering
    \includegraphics[width=\linewidth]{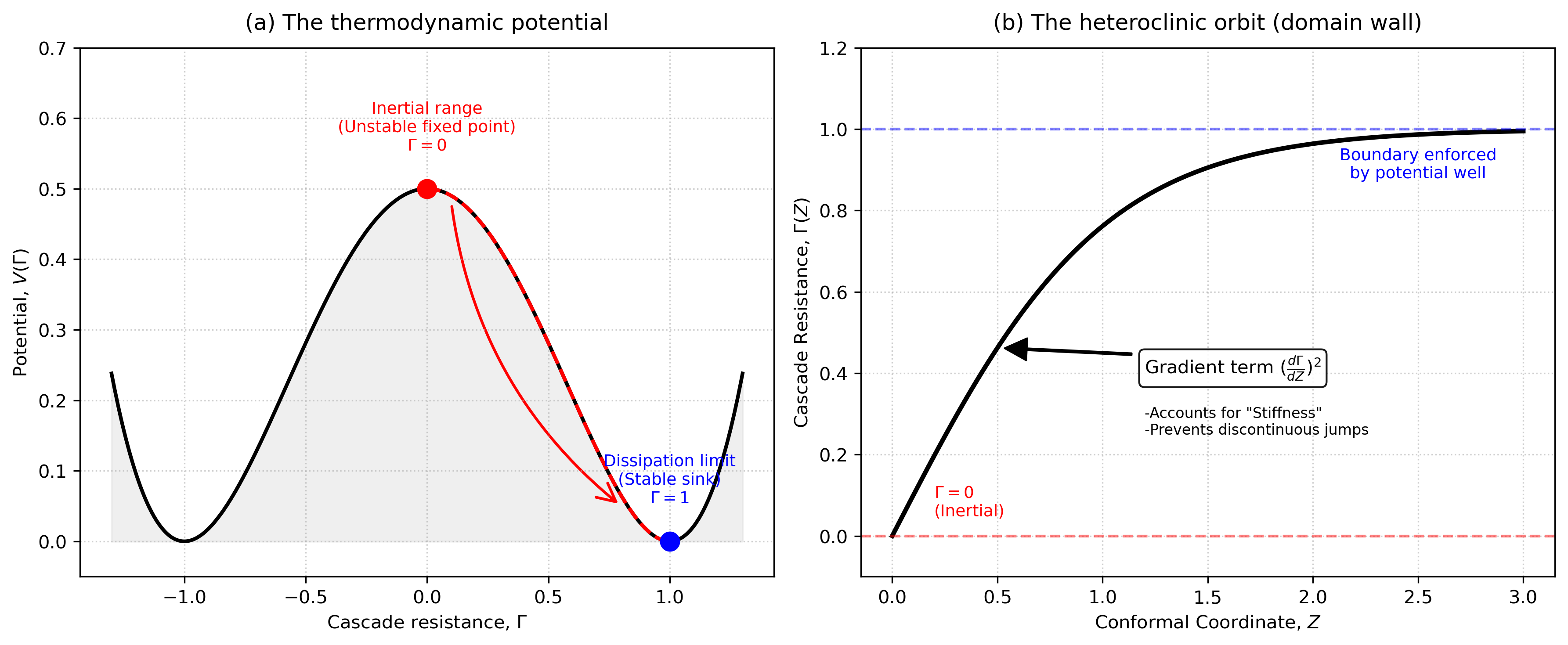}
    \caption{Schematic representation of the phenomenological Ginzburg-Landau analogy applied to cascade dynamics. \textbf{(a)} The thermodynamic potential $V(\Gamma) = \frac{1}{2}(1-\Gamma^2)^2$. The cascade dynamics follow a trajectory (solid line red arrow) from the unstable fixed point of the inertial range ($\Gamma=0$) toward the stable attractor of the dissipation limit ($\Gamma=1$), enforcing the asymptotic boundary conditions. \textbf{(b)} The resulting heteroclinic orbit solution $\Gamma(Z) = \tanh(Z)$ in conformal coordinate space. The finite slope of the transition illustrates the effect of the gradient term $(d\Gamma/dZ)^2$ in Eq. (9), which represents the spectral ``stiffness'' of the energy transfer and prevents discontinuous jumps in the cascade flux.}
    \label{fig:ginzburg_schematic}
\end{figure}

\noindent It is postulated that the steady-state cascade resistance profile minimizes a free energy functional. This functional represents a competition between spectral non-locality (penalizing abrupt changes in $\Gamma$) and the drive toward thermodynamic equilibrium (favoring complete dissipation, $\Gamma=1$):
\begin{equation}
    \mathcal{F}[\Gamma] = \int_{0}^{\infty} \left[ \frac{1}{2} \left( \frac{d\Gamma}{dZ} \right)^2 + \frac{1}{2} (1 - \Gamma^2)^2 \right] dZ
    \label{eq:free_energy}
\end{equation}
\noindent This functional adopts the structure of Ginzburg-Landau theory \cite{Goldenfeld1992}, as illustrated schematically in Figure \ref{fig:ginzburg_schematic}. In this analogy, the cascade dynamics are treated as a heteroclinic orbit connecting two fixed points: the inertial range ($\Gamma=0$) and the dissipation limit ($\Gamma=1$). The gradient term $(\frac{d\Gamma}{dZ})^2$ accounts for the ``stiffness'' of the energy transfer, preventing discontinuous jumps in spectral flux, while the double-well potential $(1-\Gamma^2)^2/2$ enforces the asymptotic boundary conditions. While phenomenological, similar variational principles have been successfully applied to derive statistical models of turbulent transport \cite{Turkington2013, Bouchet2012}.

\subsection{Euler-Lagrange solution}

\noindent Minimizing Eq.~\eqref{eq:free_energy} yields the Euler-Lagrange equation:
\begin{equation}
    \frac{d^2\Gamma}{dZ^2} = 2\Gamma(\Gamma^2 - 1)
    \label{eq:euler_lagrange}
\end{equation}

\noindent The boundary conditions $\Gamma(0) = 0$ and $\Gamma(\infty) = 1$ select the heteroclinic orbit connecting the two fixed points. The first integral of Eq.~\eqref{eq:euler_lagrange} gives:
\begin{equation}
    \frac{d\Gamma}{dZ} = 1 - \Gamma^2
    \label{eq:riccati}
\end{equation}

\noindent This Riccati equation has the well-known solution:
\begin{equation}
    \Gamma(Z) = \tanh(Z)
    \label{eq:tanh_solution}
\end{equation}

\subsection{The final spectral equation}

\noindent Substituting the conformal mapping Eq.~\eqref{eq:conformal} into Eq.~\eqref{eq:tanh_solution} and using the transmission relation Eq.~\eqref{eq:transmission_def}, the explicit form of the transmission function is obtained:
\begin{equation}
    \mathcal{T}(\kappa) = \left[1 - \tanh^2\left(\sqrt{\frac{\kappa - \kappa_d}{1 - \kappa}}\right)\right]^2
    \label{eq:transmission_final}
\end{equation}


\noindent The complete energy spectrum is:
\begin{equation}
    \boxed{E(k) = C_K \varepsilon^{2/3} k^{-5/3} \left[ 1 - \tanh^2\left(\sqrt{\frac{k\eta - k_d \eta}{1 - k\eta}}\right) \right]^2}
    \label{eq:final_spectrum}
\end{equation}

\noindent This expression is valid for $k_d\eta \leq k\eta < 1$ and is identically zero for $k\eta \geq 1$. This model is referred to as the `novel model' hereafter.

\section{Spectral reconstruction procedure}

\noindent Given an under-resolved measured spectrum $E_{meas}(k)$ for $k < k_{max}$, the reconstruction proceeds as follows.\\

\noindent \textbf{Step 1: Determine flow parameters.}
From the resolved portion of the spectrum, fit the inertial range to the Kolmogorov form $E(k) = C_K \varepsilon^{2/3} k^{-5/3}$ to obtain $\varepsilon$ \cite{Pope2000}. Calculate the Kolmogorov wavenumber $k_\eta = (\varepsilon/\nu^3)^{1/4}$ using the known kinematic viscosity.\\


\noindent \textbf{Step 2: Construct the full spectrum.}
\begin{equation}
    E_{recon}(k) = \begin{cases}
        E_{meas}(k) & k < k_{max} \\
        C_K \varepsilon^{2/3} k^{-5/3} & k_{max} \leq k < k_d\\
        C_K \varepsilon^{2/3} k^{-5/3} \mathcal{T}(k\eta) & k_d \leq k < k_\eta \\
        0 & k \geq k_\eta
    \end{cases}
    \label{eq:reconstruction}
\end{equation}
where $k_{d}$ is the wavenumber where approximately the inertial range ends and the dissipative part of the spectrum begins. Experiments have shown that $k_d\eta \approx 0.1$  \cite{Saddoughi1994,Ishihara2005}. \\

\noindent \textbf{Step 3: Integrate for TKE.}

\noindent After integration, the following is obtained: 
\begin{equation}
    \text{TKE}_{recon} = \int_0^{k_\eta} E_{recon}(k)\, dk \approx \text{TKE}
\end{equation}

\section{Results, validation, and discussion}


\noindent The model was validated against high-Reynolds-number data from Mishra \cite{mishra:tel-04908316} hot-wire experiments performed for turbulence generated using regular grids, which provides well-resolved spectra extending into the dissipation range at $Re_\lambda =\frac{u_{rms} \lambda}{U} \approx 200$--$1000$. Here, $Re_{\lambda}$ is the Taylor microscale ($\lambda$) based Reynolds number with $u_{rms}$ representing the root mean square value of the velocity fluctuations. For more information on the experiment performed, refer to chapter 5 of \cite{mishra:tel-04908316}. It should be noted that the original spectrum was obtained using a one-dimensional hot-wire probe, $E_{11}(k)$, which measures only the longitudinal component of the three-dimensional energy spectrum, $E(k)$. Since turbulence generated by a regular grid is nearly homogeneous and isotropic \cite{bailly2015homogeneous}, the full energy spectrum, $E(k)$, can be reconstructed using the relation given in Eq.~\ref{eq:relation between 3D spectrum with 2D spectrum} (see \cite{Pope2000}).

\begin{equation}
    E(k) = k^2 \frac{d^2 E_{11}(k)}{dk^2} - k \frac{d E_{11}(k)}{dk}
    \label{eq:relation between 3D spectrum with 2D spectrum}
\end{equation}


\subsection{Spectral shape comparison}

\begin{figure}[hbt!]
\centering
\includegraphics[width=\textwidth]{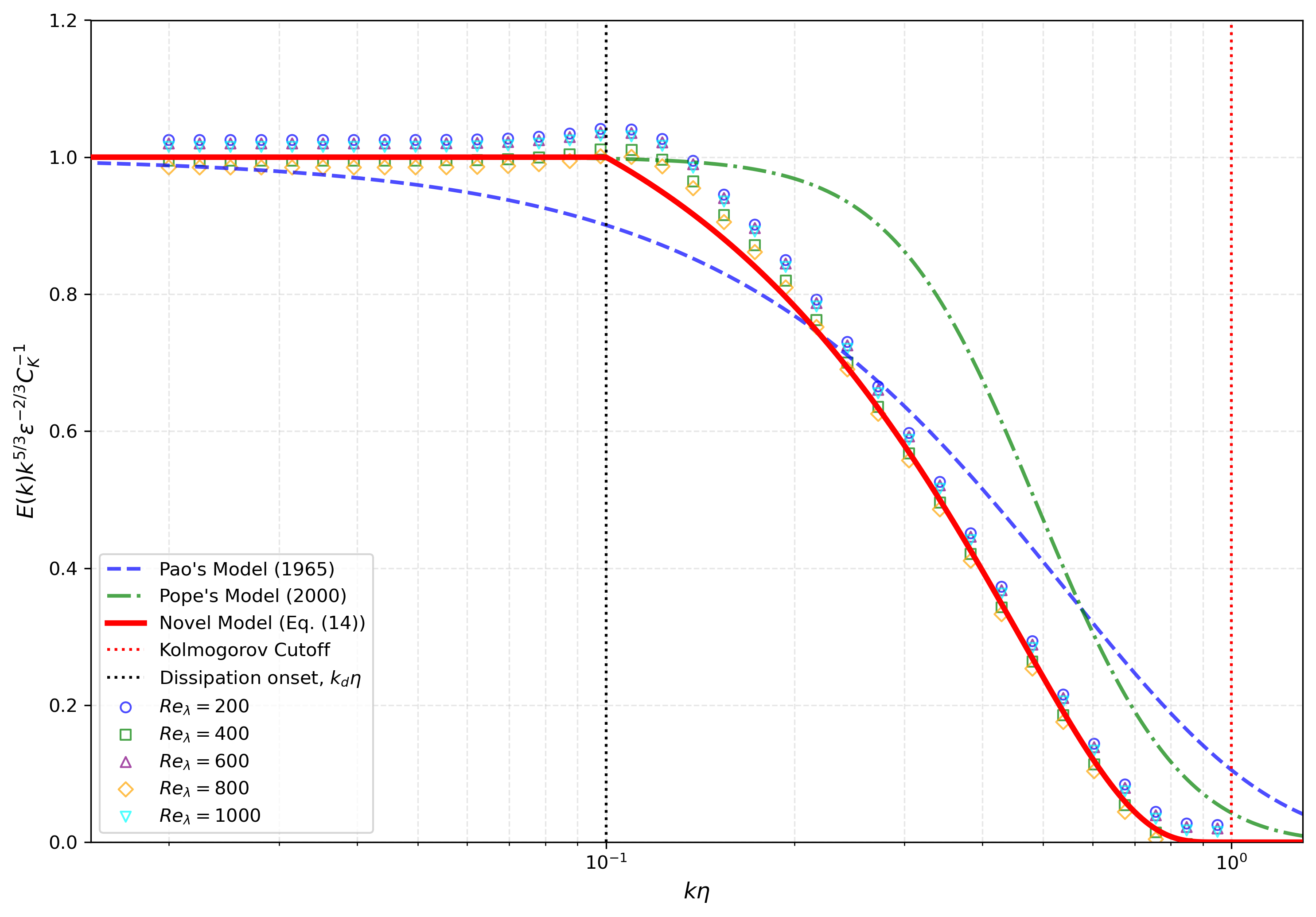}
\caption{Compensated spectrum $E(k)\varepsilon^{-2/3}k^{5/3}C_{\kappa}^{-1}$ vs $k\eta$. Comparison of novel model (Eq. (\ref{eq:final_spectrum})) (solid), Pao (dashed), Pope (dash-dot), and experimental data (symbols) from Mishra Ref.~\cite{mishra:tel-04908316}}
\label{fig:spectrum}   
\end{figure}

\noindent Figure~\ref{fig:spectrum} presents compensated energy spectra comparing the three models against experimental data from Mishra \cite{mishra:tel-04908316}. The model developed in this article clearly shows a much better match than the classical Pao \cite{Pao1965} and Pope \cite{Pope2000} models. \\ 


\begin{figure}[hbt!]
\centering
\includegraphics[width=\textwidth]{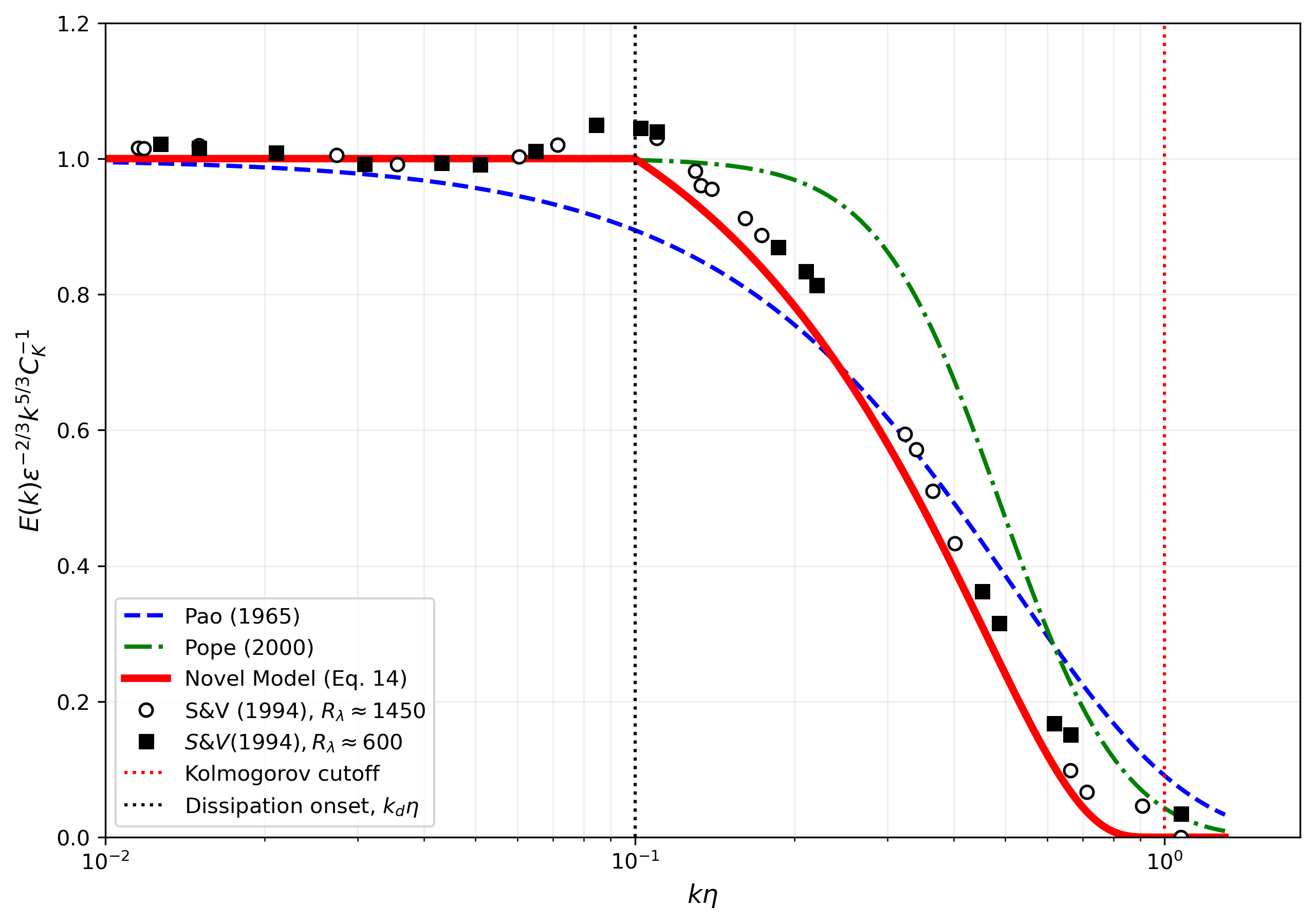}
\caption{Compensated spectrum $E(k)\varepsilon^{-2/3}k^{5/3}C_{\kappa}^{-1}$ vs $k\eta$. Comparison of novel model (Eq. (\ref{eq:final_spectrum})) (solid), Pao (dashed), Pope (dash-dot), and experimental data (symbols) from  Saddoughi and Veeravalli Ref.~\cite{Saddoughi1994}}
\label{fig:spectrum_boundary_layer}   
\end{figure}

\noindent To demonstrate the generality of the proposed model, it was compared with independent data obtained by Saddoughi and Veeravalli \cite{Saddoughi1994}. Their study utilised hot-wire anemometry to investigate a zero-pressure-gradient boundary layer at $Re_\theta \approx 370,000$. This comparison is presented in figure~\ref{fig:spectrum_boundary_layer}. The proposed model (Eq.~\ref{eq:final_spectrum}) demonstrates good agreement with the turbulence data from these boundary layer measurements.\\

\noindent The consistency of the model (Eq.~\ref{eq:spectrum}) with both regular grid-generated turbulence \cite{mishra:tel-04908316} and boundary-layer turbulence \cite{Saddoughi1994} confirms its robustness across different flow regimes. This validates the use of the model for reconstructing the dissipative range of the energy spectrum in cases where experimental limitations prevent full resolution.

\subsection{TKE recovery comparison}

\noindent To simulate under-resolved measurements, the fully-resolved spectra were artificially truncated at various $k_{max}$ values for the $Re_{\lambda} = 800$ case of Mishra \cite{mishra:tel-04908316}. The truncated spectra were then reconstructed using: (a) the novel model (Eq. \ref{eq:final_spectrum}), (b) the Pao model \cite{Pao1965}, and (c) the Pope model \cite{Pope2000}. Since the models proposed by Pao \cite{Pao1965} and Pope \cite{Pope2000} feature exponential tails, they have been truncated at $k\eta = 1$ to facilitate a consistent comparison. TKE was computed by integration and compared to the ground-truth value from the full spectrum. Table~\ref{tab:validation} presents the TKE recovery performance for the three models at various truncation wavenumbers. The present model consistently achieves higher accuracy than both the Pao and Pope formulations across all truncation levels tested, where Pao's model \cite{Pao1965} underestimates TKE, and Pope's model \cite{Pope2000} overestimates it.


\begin{table}[h]
\centering
\caption{TKE recovery from truncated spectra (\% of true TKE)}
\label{tab:validation}
\begin{tabular}{@{}lcccc@{}}
\toprule
Truncation $k_{max}\eta$ & Truncated & Novel Model (Eq. \ref{eq:final_spectrum}) & Pao & Pope \\
\midrule
0.05 & 89.2 & 98.7 & 94.3 & 105.1 \\
0.10 & 93.1 & 99.2 & 96.4 & 107.0 \\
0.15 & 95.4 & 99.5 & 97.6 & 112.1 \\
0.20 & 96.8 & 99.1 & 98.2 & 115.5 \\
\bottomrule
\end{tabular}
\end{table}




    

\subsection{Advantages of the compact support formulation}

\noindent The formulation of a spectral cutoff (compact support) at the Kolmogorov scale offers distinct methodological advantages for TKE reconstruction, particularly when compared to asymptotic decay models:

\begin{enumerate}
    \item \textbf{Energetic sufficiency vs. asymptotic decay:} While strict solutions to the Navier-Stokes equations imply asymptotic spectral decay due to viscous diffusion \cite{Tennekes1972}, the energy density contained in the far-dissipation range ($k\eta > 1$) is negligible for integrated quantities. The proposed cutoff acts as an effective energetic closure, capturing the totality of the cascade flux while avoiding the modelling uncertainties associated with far-dissipation intermittency. By treating the Kolmogorov scale as an effective sink, the model recovers over 99\% of the variance without requiring the resolution of exponential tails.

    \item \textbf{Well-posed spectral integration:} Classical models (e.g., Pao \cite{Pao1965}, Pope \cite{Pope2000}) extend to $k \rightarrow \infty$, necessitating arbitrary truncation limits during numerical integration to avoid diverging errors or numerical noise. The present model features bounded spectral support $\kappa \in [\kappa_d, 1]$, rendering the TKE integration definitive and unambiguous. This eliminates user-dependent bias in determining the ``end'' of the spectrum.

    \item \textbf{Parameter universality:} The reconstruction relies solely on the Kolmogorov constant $C_K$. Unlike exponential-tail models, which often require fitting parameters ($\beta, c_{\eta}$) that exhibit weak Reynolds-number dependence \cite{Martinez1997, Ishihara2005}, the present formulation is parameter-free in the dissipation range. This makes it particularly robust for reconstruction, where the exact Reynolds number of the flow may be unknown or fluctuating.
\end{enumerate}
    

\subsection{Limitations}

\noindent The method assumes locally isotropic turbulence, consistent with Kolmogorov's hypotheses \cite{Kolmogorov1941}. In flows with strong anisotropy extending into the dissipation range, such as near-wall turbulence or highly strained flows, deviations from the model may occur \cite{Antonia1991,Sreenivasan1998}.

\noindent The onset wavenumber $\kappa_d = 0.1$ is empirically determined from the observed extent of the inertial range in high-Reynolds-number flows \cite{Saddoughi1994}. While this value is approximately universal for $Re_\lambda \gtrsim 100$ \cite{Pope2000}, weak Reynolds-number dependence may exist at lower $Re_\lambda$ \cite{Martinez1997}.

\subsection{Recommended applications}

\noindent The reconstruction method is particularly suited for:
\begin{itemize}
     \item Regular grid generated turbulence characterization in a wind tunnel using pressure rakes or surface-mounted transducers \cite{Blake2017}
    \item Aeroacoustic testing where pressure sensors are used to characterize inflow turbulence \cite{Amiet1975}
    \item Industrial flow diagnostics where hot-wire deployment is impractical
    \item Atmospheric turbulence measurements with sonic anemometers, which have limited bandwidth at small scales \cite{Kaimal1994}

\end{itemize}

\section{Conclusion}

\noindent A spectral reconstruction method has been presented for estimating turbulent kinetic energy from under-resolved measurements. The method employs an analytical dissipation-range model derived from variational principles, yielding a Ginzburg-Landau domain wall solution \cite{Goldenfeld1992} with a hard cutoff at the Kolmogorov wavenumber.

\noindent The consistency of the model (Eq.~\ref{eq:spectrum}) with both homogeneous grid-generated turbulence \cite{mishra:tel-04908316} and high-Reynolds-number boundary layers \cite{Saddoughi1994} confirms its robustness across distinct flow regimes. Crucially, this agreement is achieved without flow-specific calibration parameters. This validates the model as a generalizable tool for reconstructing the dissipative range in cases where experimental limitations prevent full spectral resolution. Also, validation against experimental data demonstrates TKE recovery exceeding 98\% from spectra truncated at $k\eta = 0.15$, outperforming both the Pao \cite{Pao1965} and Pope \cite{Pope2000} exponential-tail models in terms of accuracy. It also demonstrated that the parameter-free formulation and bounded spectral support make the method well-suited for practical turbulence characterization using bandwidth-limited sensors.

\section*{Acknowledgments}

\noindent The author would like to express sincere gratitude to Dr.\ Caroline Braud of the CNRS, whose question---``Is it possible to develop a methodology that can help the aerodynamics community determine the inflow turbulence intensity to a reasonably accurate value when the sensors involved are unable to resolve the full turbulence spectrum?''---prompted the development of this work and manuscript. The author also gratefully acknowledges the financial support provided by CARNOT MER's LIFEMONITOR project.

\bibliography{references}

\end{document}